\newcommand{\nc}{\newcommand}
\nc{\beq}{\begin{equation}}   \nc{\eeq}{\end{equation}}
\nc{\bea}{\begin{eqnarray}}   \nc{\eea}{\end{eqnarray}}
\nc{\baa}{\begin{array}}      \nc{\eaa}{\end{array}}
\nc{\bit}{\begin{itemize}}    \nc{\eit}{\end{itemize}}
\nc{\ben}{\begin{enumerate}}  \nc{\een}{\end{enumerate}}
\nc{\bce}{\begin{center}}     \nc{\ece}{\end{center}}
\nc{\ufa}{\mbox{\tiny (-$\!$-$\!$-)}}
\nc{\opa}{\!}
\def\beqa{\begin{eqnarray}}
\def\eeqa{\end{eqnarray}}

\def\lsim{\mathrel{\raise.3ex\hbox{$<$\kern-.75em\lower1ex\hbox{$\sim$}}}}
\def\gsim{\mathrel{\raise.3ex\hbox{$>$\kern-.75em\lower1ex\hbox{$\sim$}}}}

\def\ie{{\it i.e. }}

\def\be{\beq}
\def\ee{\eeq}
\def\to{\rightarrow}

\def\PHYSICA #1 #2 #3 {{\sl Physica}~{\bf#1} (#3) #2}
\def\MPL #1 #2 #3 {{\sl Mod.~Phys.~Lett.}~{\bf#1} (#3) #2}
\def\NPB #1 #2 #3 {{\sl Nucl.~Phys.}~{\bf #1} (#3) #2}
\def\NPBPS #1 #2 #3 {{\sl Nucl.~Phys.~B~(Proc. Suppl.)}~{\bf #1} (#3) #2}
\def\PLB #1 #2 #3 {{\sl Phys.~Lett.}~{\bf #1} (#3) #2}
\def\PR #1 #2 #3 {{\sl Phys.~Rep.}~{\bf#1} (#3) #2}
\def\PRD #1 #2 #3 {{\sl Phys.~Rev.}~{\bf #1} (#3) #2}
\def\PRL #1 #2 #3 {{\sl Phys.~Rev.~Lett.}~{\bf#1} (#3) #2}
\def\RMP #1 #2 #3 {{\sl Rev.~Mod.~Phys.}~{\bf#1} (#3) #2}
\def\ZPC #1 #2 #3 {{\sl Z.~Phys.}~{\bf #1} (#3) #2}
\def\IJMP #1 #2 #3 {{\sl Int.~J.~Mod.~Phys.}~{\bf#1} (#3) #2}

\documentstyle[preprint,epsfig,epsf,aps,floats]{revtex}
\begin{document}  
\newlength{\captsize} \let\captsize=\small 
\newlength{\captwidth}                     

\tightenlines

\title{Quantum dissipative effects and neutrinos : current constraints and 
future perspectives}

\preprint{\vbox{\hbox{IFUSP-DFN/055-2000}
\hbox{hep-ph/0009222}
}}

\author{A.\ M.\ Gago$^{1,2}$~\thanks{Email address: agago@charme.if.usp.br},
E.\ M.\ Santos$^{1}$~\thanks{Email address: emoura@charme.if.usp.br},
W.\ J.\ C.\ Teves$^{1}$~\thanks{Email address: teves@charme.if.usp.br} and 
R.\ Zukanovich Funchal$^{1}$~\thanks{Email address: zukanov@charme.if.usp.br}\\}

\vspace{1.cm}

\address{\sl 
$^1$ Instituto de F\'\i sica, Universidade de S\~ao Paulo,
    C.\ P.\ 66.318, 05315-970\\  S\~ao Paulo, Brazil\\
$^2$ Secci\'on F\'{\i}sica, Departamento de Ciencias,
    Pontificia Universidad Cat\'{o}lica del Per\'{u}\\ 
    Apartado 1761, Lima, Per\'{u}\\
\vglue -0.2cm
}

\maketitle
\vspace{.5cm}
\hfuzz=25pt
\begin{abstract}
\noindent 
We establish the most stringent experimental constraints coming 
from recent terrestrial neutrino experiments on quantum mechanical 
decoherence effects in neutrino systems.
Taking a completely phenomenological approach, we probe vacuum 
oscillations plus quantum decoherence between two neutrino species in the 
channels $\nu_\mu \to \nu_\tau$, $\nu_\mu \to \nu_e$ and 
$\nu_e \to \nu_\tau$, admitting that the quantum decoherence 
parameter $\gamma$ is related to the neutrino energy $E_\nu$ 
as : $\gamma=\gamma_0 \, E_{\nu}^{n}$, with $n=-1,0,1$ and $2$.
Our bounds are  valid for a neutrino mass squared 
difference compatible with the atmospheric, the solar and, 
in many cases, the LSND scale.
We also qualitatively discuss the perspectives of the future long baseline 
neutrino experiments to further probe quantum dissipation. 
\end{abstract}
\pacs{PACS numbers: }

\section{Introduction}
\label{sec:intro}

The striking results of solar~\cite{solar} and atmospheric~\cite{atmos} 
neutrino experiments testify, beyond any reasonable doubt, that neutrino 
physics involves quantum interference phenomena. 
This is why it is plausible  to envisage today the use of neutrino 
oscillations to probe the foundations of  quantum mechanics and, in 
particularly, to test the completeness of the theory.

From the theoretical point of view, quantum dissipative 
effects can be viewed either as a consequence of a fundamental violation of  
quantum physics~\cite{hawk1,hawk2}, motivated by quantum gravity,
or of an effective description of a micro system weakly coupled to the 
macroscopic world in a open quantum system framework~\cite{open}.

In the former approach, it is argued that quantum fluctuations of the 
gravitational field can lead to a loss of quantum coherence, 
making time-evolution transform pure quantum states into mixed ones, thereby 
violating ordinary quantum mechanics~\cite{ellis1,sc,ljg}. In the latter 
approach, quantum mechanics is not violated at the level of the global system  
but rather only by the reduced effective dynamics of the micro subsystem 
weakly coupled to the macro reservoir~\cite{open,fb}.

Since both of these attitudes, although conceptually very different, will 
result in a modification of the time evolution of the neutrino mass 
eigenstates leading to the appearance of damping factors in the neutrino 
oscillation probabilities, we will not advocate in favor of either physical 
interpretation but rather simply treat here the effect phenomenologically. 
We consider that the absence of a full dynamical theory that 
can account for the origin, define the energy dependence and ultimately 
estimate the size of the decoherence effect is a good motivation for  
phenomenological analyses which can derive, directly from experimental data, 
limits which are valid independently of the (unknown) theoretical picture.

Bounds on dissipative parameters have already been derived from 
studies of neutral meson systems~\cite{open,jmn,cplear,hp,meson} and neutron 
interferometry~\cite{neutron}. A first attempt to use  neutrino systems 
to investigate quantum dissipative effects was reported 
in Ref.~\cite{chinese0}. In Ref.~\cite{chinese1}, the possibilities of 
future neutrino experiments to unravel quantum decoherence was 
qualitatively discussed. Recently, tight limits were obtained in 
the channel $\nu_\mu \to \nu_\tau$ from the atmospheric neutrino data, 
for a decoherence parameter $\gamma$ which is either supposed to be a 
constant or proportional to the neutrino energy squared, and in fact a 
solution to the atmospheric neutrino problem (ANP) was found, when $\gamma$ 
was suppose to be inversely proportional to the neutrino energy~\cite{lisi}. 
The quantum decoherence parameter at the best-fit value of this novel 
solution to the ANP is such that it is not in conflict with other 
experimental constraints, but  it is big enough to explain all the 
atmospheric neutrino data (sub-GeV, multi-GeV and upward going muons) 
comparably well as the mass induced $\nu_\mu \to \nu_\tau$ oscillation 
mechanism in vacuum.

In this paper, we investigate the most stringent constraints on the 
quantum dissipation parameters one can get from terrestrial neutrino 
experiments, considering flavor conversions between only two neutrino species 
and assuming that these parameters have some specific energy dependence. 
We will suppose that, in nature, mass induced vacuum oscillations are 
accompanied by quantum decoherence.  In this framework, we will extract 
our limits on decoherence using data from experiments that 
have not registered any signal of flavor conversion,  so that our bounds 
will be valid from a maximal value of $\Delta m^2$, which will depend on each 
experimental setup, down to $\Delta m^2=0$. 
Although our bounds are more general, they will be true, in particular, 
in the range of  $\Delta m^2$ consistent with the $\nu_\mu \to \nu_\tau$ 
oscillation solution to the ANP, 
\ie $\Delta m^2 \sim (2-5)\times 10^{-3}$~eV$^2$~\cite{atm-ana} and the 
$\nu_e \to \nu_\mu, \nu_\tau$ solutions to the solar neutrino problem (SNP)  
either in vacuum, with $\Delta m^2 \sim 10^{-8} \text{ eV}^2-10^{-11}$~eV$^2$~\cite{vac-ana}, or in matter through the MSW mechanisms, with 
$\Delta m^2 \sim (2-20) \times 10^{-5}$~eV$^2$ (LMA) or 
$\Delta m^2 \sim (4-10) \times 10^{-6}$~eV$^2$ (SMA) 
or $\Delta m^2 \sim (6-20) \times 10^{-8}$~eV$^2$ (LOW)~\cite{msw-ana}. 
In the majority of cases for $\nu_\mu \to \nu_e$, our constraints can also 
be applied in the LSND allowed region, 
$3 \times 10^{-3} \leq  \sin^2 2\theta \leq 3 \times 10^{-2}$ and 
$0.2 \text{ eV}^2 \; \lsim \Delta m^2 \lsim 2 $ 
eV$^2$~\cite{lsnd,karmen,bugey}.
We will stress our results in these $\Delta m^2$ regions, since 
they are, from the point of view of  evidences in favor of flavor 
change, the most interesting ones.
Moreover, we have to emphasize that in the point $\sin^2 2\theta=1$, our 
limits  can be re-interpreted as a bound on flavor conversion driven by 
the pure decoherence mechanism alone ($\Delta m^2=0$).
 
After establishing the best current constraints, we briefly discuss the 
capability of future long baseline neutrino experiments to expand these 
limits. 

The outline of the paper is as follows.
In Sec.\ \ref{sec:sec2}, we review  how to introduce quantum 
decoherence in the evolution equation of the neutrino mass eigenstates in 
the density matrix formalism. 
Using this modified formalism, we introduce the quantum decoherence  
parameters and justify the form of the neutrino oscillation probability 
we will use in this work.
In Sec.\ \ref{sec:sec3}, we describe how we have analyzed the experimental 
data  from CCFR~\cite{ccfr}, E776~\cite{e776}, CHORUS~\cite{chorus,chorus1,chorus2,chorus3,chorus4} and CHOOZ~\cite{chooz} in order to extract our limits.
In Sec.\ \ref{sec:sec4}, we discuss the most important 
constraints from terrestrial experiments on the quantum decoherence 
parameters in the neutrino oscillation channels $\nu_\mu \to \nu_\tau$, 
$\nu_\mu \to \nu_e$ and $\nu_e \to \nu_\tau$. In Sec.\ \ref{sec:sec5}, we 
argue on the perspectives of  KamLAND and other future 
neutrino experiments to put even more restrictive bounds on quantum 
dissipation. Finally, in Sec.\ \ref{sec:sec6}, we present our conclusions.

\section{Review of the Formalism} 
\label{sec:sec2}

The time evolution of neutrinos created in a given flavor $\nu_\alpha$ 
by  weak interactions, as of any quantum state,  can be described 
using  the density matrix formalism by the Liouville equation.
In this formalism, the neutrino state in time can be described by a 
density matrix $\rho_{\alpha}$,  which is a hermitian operator, with 
positive eigenvalues and constant trace. 
One can suppose that $\rho_{\alpha}$ also has the additional 
property of being completely positive according to Refs.~\cite{fb,cp}, but 
this important theoretical point will not be crucial in the limit we are 
interested here. We will comment more on this  below. 

Considering two neutrino generations, in the basis of the two mass 
eigenstates $\nu_1$ and $\nu_2$ that have masses $m_1$ and $m_2$, 
respectively, the two flavor eigenstates $\nu_\alpha$ and $\nu_\beta$ 
can be represented by $2\times2$ matrices as 

\begin{equation}
\rho_{\alpha}= \left( \begin{array}{cc}
\cos^2 \theta &   \cos \theta \sin \theta  \\ 
\cos \theta \sin \theta  & \sin^2 \theta   
\end{array} 
\right), 
\label{nue}
\end{equation}
and

\begin{equation}
\rho_{\beta}= \left( \begin{array}{cc}
\sin^2 \theta &  - \cos \theta \sin \theta  \\ 
-\cos \theta \sin \theta  & \cos^2 \theta   
\end{array} 
\right) \equiv 1 - \rho_{\alpha}, 
\label{numu}
\end{equation}
where $\theta$ is the usual Cabibbo-like mixing angle that parametrizes the 
matrix which relates mass and flavor eigenstates.

If we add an extra term $L[\rho_{\alpha}]$ to the Liouville equation,  
quantum states can develop dissipation and irreversibility~\cite{ellis1,fb}. 
The generalized Liouville equation for $\rho_{\alpha}(t)$ can then be written 
as~\cite{fb}

\begin{equation}
\partial_t \rho_{\alpha}(t)= -\imath [H,\rho_{\alpha}(t)] +L[\rho_{\alpha}(t)], \label{evol}
\end{equation} 
where the effective hamiltonian $H$ is, in vacuum, given by 
\begin{equation}
H= \left[ \begin{array}{cc}
 \Delta & 0 \\ 
0 & -\Delta 
\end{array} 
\right], 
\label{ham}
\end{equation} 
where $\Delta = (m^2_2-m^2_1)/4E_\nu$, we have already considered 
ultra-relativistic neutrinos of energy $E_\nu$ and the irrelevant global 
phase has been subtracted out.

One can rewrite Eq.\ (\ref{evol}) using as basis the identity and the Pauli 
matrices ($\sigma_\mu$, $\mu =0,1,2,3$), here we drop the neutrino flavor 
index for simplicity,

\begin{equation}
\partial_t \rho_\mu \, \sigma_\mu = 2 \epsilon_{\mu \nu \delta} h_\mu \rho_\nu \sigma_\delta +L_{\mu \nu} \rho_\nu \sigma_\mu,
\label{expan}
\end{equation} 
where sum over repeated indices is implied, 
$ H = h_\mu \sigma_\mu$, $\rho = \rho_\mu \sigma_\mu$ and   

\begin{equation}
L_{\mu \nu} = -2 \left[ \begin{array}{cccc}
0 & 0 & 0 & 0 \\ 
0 & a & b & c \\ 
0 & b & \gamma & \beta \\ 
0 & c & \beta & \alpha  
\end{array} 
\right], 
\label{luv}
\end{equation}
is the most general parametrization for $L[\rho]$, it contains six real 
parameters which are not independent if one assumes the complete positivity 
condition~\cite{cp}. While the requirement of simple positivity guarantees 
that the eigenvalues of $\rho$, interpreted as probabilities in the formalism, 
remain positive at any time, the requirement of complete positivity guarantees 
the same thing for the density matrix which describes the larger system 
(neutrino plus environment). This assures the absence of unphysical effects, 
such as negative probabilities, when dealing with correlated 
systems~\cite{fb,cp}. The relations that must satisfy these parameters, if one 
assumes complete positivity, can be found in Ref.~\cite{fb}. 
 
Using the above parametrization, Eq.\ (\ref{expan}) can be written as 
\begin{mathletters}
\label{equas}
\begin{eqnarray}
\partial_t \rho_0(t) & = & 0, \\
\partial_t \rho_1(t) & = & 2 \epsilon_{\mu \nu 1} h_\mu \rho_\nu - 2 [ a \rho_1 + b \rho_2 + c \rho_3],  \\
\partial_t \rho_2(t)  &= &2 \epsilon_{\mu \nu 2} h_\mu \rho_\nu - 2 [ b \rho_1 + \gamma \rho_2 + \beta \rho_3],  \\
\partial_t \rho_3(t)  &= &2 \epsilon_{\mu \nu 3} h_\mu \rho_\nu - 2 [ c \rho_1 + \beta \rho_2 + \alpha \rho_3].
\end{eqnarray}
\end{mathletters}

These equations lead, in general, to a $\nu_\alpha \to \nu_\beta$ transition 
probability that will present a time behavior characterized by two kinds of 
contributions: an oscillating one, controlled by $\Delta$, and an 
exponentially damping one, driven by the dissipative parameters. 
The importance of these effects will depend on the relative magnitude of 
$\Delta$ with respect to the decoherence parameters. In particular, an 
asymptotically oscillatory behavior will be possible if these decoherence 
parameters are all small when compared to $\Delta$. 
One can also envisage, as an opposite extreme case, the situation where 
decoherence is the dominant phenomenon. In this case there will be no 
oscillation but only a damping effect. 

Here we are interested in investigating the possibility of extracting limits 
on decoherence admitting that neutrino oscillation is the dominant process.  
The simplest situation is the one in which we can have the oscillation 
probability in the presence of a single extra dissipation factor. 
In fact, this situation physically arises  when the weak coupling 
limit condition is satisfied, and corresponds 
to $\alpha=0 \rightarrow a=\gamma, b=c=\beta=0$, if we impose complete 
positivity. The weak coupling limit is one way of implementing the physical 
requirement that the interaction between neutrinos and environment is weak, 
as explained in Ref.~\cite{fb}.
Even if complete positivity is not assume, it is reasonable to think that 
all these parameters should take very small values, otherwise 
decoherence would be a predominant effect, so that the simplest situation 
would still be to consider that only one of them is large enough to be 
within experimental reach. In this case, Eqs.\ (\ref{equas}) simplify to 
\begin{mathletters}
\label{eqss}
\begin{eqnarray}
\rho_0(t)  &= & \rho_0(0),  \\
\partial_t \rho_1(t) & = &2 [h_2 \rho_3 - h_3 \rho_2] - 2  \gamma \rho_1,  \\
\partial_t \rho_2(t) & = &2 [h_3 \rho_1 - h_1 \rho_3] - 2 \gamma \rho_2,  \\
\partial_t \rho_3(t) & = &2 [h_1 \rho_2 - h_2 \rho_1].
\end{eqnarray}
\end{mathletters}
Solving Eqs.\ (\ref{eqss}), with  $h_1=h_2=0$ and $h_3=-\Delta$ from 
Eq.\ (\ref{ham}) and the initial conditions which correspond to 
$\rho_\alpha(0)=\rho_\alpha$, 
\begin{mathletters}
\label{solucao}
\begin{eqnarray}
\rho_0(t) = & \frac{1}{2}, \\
\rho_1(t) = & \frac{1}{2} \sin 2 \theta \;e^{-2\gamma t} \cos ( 2 \Delta t), \\
\rho_2(t) = & -\frac{1}{2} \sin 2 \theta \;e^{-2\gamma t} \sin ( 2 \Delta t), \\
\rho_3(t) = & \frac{1}{2} \cos 2 \theta,
\end{eqnarray}
\end{mathletters}
so that the flavor conversion probability, which in any case can be calculated 
as 
\begin{eqnarray}
P(\nu_\alpha \to \nu_\beta) (t) &=& 
\text{Tr}[\rho_{\alpha}(t) \rho_{\beta}(0)] \nonumber \\
&=& 1-[\frac{1}{2}+\rho_3(t)]\cos^2 \theta - [\frac{1}{2}-\rho_3(t)]\sin^2 \theta - \rho_1(t) \sin 2\theta, 
\label{prob}
\end{eqnarray} 
gives in the simplest, but physically meaningful, limit we are 
interested in  
\begin{equation}
P(\nu_\alpha \to \nu_\beta) = \frac{1}{2} \sin^2 2 \theta \; [1- e^{-2 \gamma L} \cos (2 \Delta L)],
\label{proba}
\end{equation} 
the probability of finding the neutrino produced in the flavor 
state $\nu_\alpha$ in the flavor state $\nu_\beta$ after traveling a 
distance $L$ under the influence of quantum dissipation driven by the parameter $\gamma$.  If $\gamma=0$, we recover the usual mass induced oscillation 
probability in two generations.
Although other possibilities exist, where all decoherence parameters 
survive, this simple form will at least allow us to obtain from 
experimental data a very good idea of the order of magnitude of limits 
we can get on decoherence parameters.

Notice that, by Eq.\ (\ref{prob}), even if there is no mixture, \ie 
$\cos \theta =1$ and $\sin \theta =0$, there can be transition 
as long as $\rho_3(t)$ is not a constant, so that even if neutrinos are 
degenerate or massless, flavor change can still take place through the 
pure decoherence mechanism (PDM). The simplest case of flavor conversion via 
PDM ($\Delta m^2=0$) is the one where only a single decoherence parameter that 
appears in Eq.~(7d), say $\alpha=\gamma$, is non zero. This implies  
the probability of conversion

\begin{equation}
P(\nu_\alpha \to \nu_\beta) = \frac{1}{2} \; [1- e^{-2 \gamma L}].
\label{pdm}
\end{equation}

\section{Description of the Experimental Data Analyses} 
\label{sec:sec3}

We analyze data from the neutrino accelerator experiments CHORUS, CCFR, 
and E776, and the reactor experiment CHOOZ; which can currently put the 
 most stringent limits on the dissipation parameter $\gamma$, in the 
 context of two neutrino generations. The exact dependence of the dissipation 
 parameter $\gamma$ with the neutrino energy is unknown.
 Several forms, based on physical considerations, 
have been proposed in the literature~\cite{ellis1,fb,chinese1,lisi}. 
For this reason, we will investigate in this work four different 
{\em ansatz}  on the $\gamma$ energy dependence:

\begin{equation}   
\gamma = \gamma_0 \, E_{\nu}^{n},  
\; \; \; \; n=-1,0,1 \text{ and } 2\,, 
\label{gam}
\end{equation}
where $E_\nu$ is the neutrino energy in GeV and $\gamma_0$ is a constant 
given in units of GeV$^{(1-n)}$.

In the case of the specific baseline and energy range of the investigated 
experiments, with the mass squared differences that we will consider here, 
the cosine  term in Eq.\ (\ref{proba}) turns out  to be 1, so that one can 
simply write

\begin{equation}
P(\nu_\alpha \to \nu_\beta) = \frac{1}{2} \sin^2 2 \theta \; 
[1- e^{-2 \gamma L} ], 
\label{simpro}
\end{equation}
where $\gamma$ is given by Eq.\ (\ref{gam}) and 
$(\alpha,\beta)=(\mu,\tau), (\mu,e)$, $(e,\tau)$. The survival 
probability $P(\nu_\alpha \to \nu_\alpha)=1-P(\nu_\alpha \to \nu_\beta)$, 
in any given channel, merely by conservation of probability. 
Note that, for $\sin^2 2 \theta=1$, the above formula is 
 equal to the one for the PDM given in Eq.\ (\ref{pdm}), so 
that at this particular point, our results can also be interpreted as a 
direct limit on pure decoherence. In spite of that, the two situations are 
not physically equivalent. 

From Eq.\ (\ref{simpro}), one can have a quick idea about the bound on 
$\gamma_0$ that can be reached by a given experiment. Due to 
the fact that we are dealing
 with experiments that give no evidence of neutrino conversion, the condition
 $P(\nu_\alpha \to \nu_\beta) < P_{\text{min}}$ must be satisfied, where 
 $P_{\text{min}}$ is the minimum probability (or sensitivity) that could be
 measured at each experiment. This means that 
\be
[1- e^{-2 \gamma L} ] < \displaystyle \frac{2 \,P_{\text{min}}}{\sin^2 2 \theta},
\label{pmaxlim}
\ee
thus
\begin{eqnarray}
\gamma_0 & \lsim & -\,\frac{\ln{\left(1-2 \displaystyle
\frac{P_{\text{min}}}{\sin^2 2\theta}\right)}}
{(L/\text{km}) \,<E_\nu/\text{GeV} >^n} \times 10^{-19}\text{ GeV}^{(1-n)},  
\label{estimation}
\end{eqnarray}
where $n=-1,0,1$ and $2$, $L$ is the neutrino flight length in km and $<E_\nu>$ the neutrino mean energy in GeV. This equation permits us to know in 
advance which experiment will put the best bound on $\gamma_0$ for a 
given $n$, in each oscillation mode.
 
We derive constraints on $\gamma_0$ from three oscillation channels,  
 $\nu_\mu \to \nu_\tau$, $\nu_\mu \to \nu_e$ and $\nu_e \to \nu_\tau$:
\begin{enumerate}
\item For $\nu_\mu \to \nu_\tau$, we focus our attention in the 
range of $\sin^2 2 \theta$ consistent with ANP, 
\ie $0.8 \leq  \sin^2 2\theta \leq 1.0$. In all the cases of $n$,  
 we work with CHORUS, our constraints being valid 
for  $\Delta m^2 \lsim 0.4$ eV$^2$.
\item For $\nu_\mu \to \nu_e$, we choose to stress two different regions of
  $\sin^2 2 \theta$: $0.5 \leq  \sin^2 2\theta \leq 1.0$ and
 $6 \times 10^{-3} \leq  \sin^2 2\theta \leq 6 \times 10^{-2}$.
 The first, we will call the large amplitude case (LA) and the second
 the small amplitude case (SA). Our interest in these regions is 
because they cover a similar range in $\sin^2 2 \theta$  
as the standard oscillation solutions to the SNP~\footnote{Note that since we are working in two generations $P(\nu_\mu \to \nu_e) \equiv P(\nu_e \to \nu_\mu)$ by T invariance.}. We remark that, LA encompasses the vacuum, the MSW large 
and the MSW low mixing angle solutions, while SA includes a big part of the 
MSW small mixing angle solution to the SNP and of the region allowed by 
LSND~\cite{lsnd}, KARMEN~\cite{karmen} and Bugey~\cite{bugey} experiments. 
For $n=-1$, we obtain the best constraints on $\gamma_0$ with CHOOZ for LA, 
and E776 for SA, valid from $\Delta m^2 \lsim 1 \times 10^{-3}$~eV$^2$ and 
  $\Delta m^2 \lsim 5 \times 10^{-2}$~eV$^2$, respectively.
For $n=0,1,2$, CCFR gives the best limits in both cases, LA and SA, which  
are all valid  from $\Delta m^2 \lsim 2$ eV$^2$.
\item For $\nu_e \to \nu_\tau$, we emphasize only the LA case, based     
on the same motivations explained for  the 
$\nu_\mu \to \nu_e$ mode. We get the best bounds on  
$\gamma_0$ for $n=-1$ with CHOOZ~($\Delta m^2 \lsim 1 \times 10^{-3}$~eV$^2$),
 for $n=0,1$ with CHORUS ($\Delta m^2 \lsim 4$~eV$^2$), and for $n=2$ with 
CCFR ($\Delta m^2 \lsim 20 $~eV $^2$). 
\end{enumerate}

At the following sections, we will describe the  
data analysis performed  for each experiment in order to calculate 
limits on $\gamma_0$ with a precise statistical significance. 

\subsection{CCFR}
\label{sec:sec3a}
The CCFR experiment at Fermilab~\cite{ccfr} is a neutrino oscillation 
appearance experiment which provides stringent limits on 
three different oscillation channels: (1) $\nu_\mu \to \nu_\tau$, 
(2) $\nu_\mu \to \nu_e$ and (3) $\nu_e \to \nu_\tau$.
The neutrino beam in this facility consists of about 
98 \% $(\nu_\mu+\bar \nu_\mu)$ and 2 \% $(\nu_e+\bar \nu_e)$.  
The experiment has a 1.4 km oscillation length of which 0.5 km is the decay 
region. 

\begin{figure}
\vglue -4.0cm
\centerline{
\epsfig{file=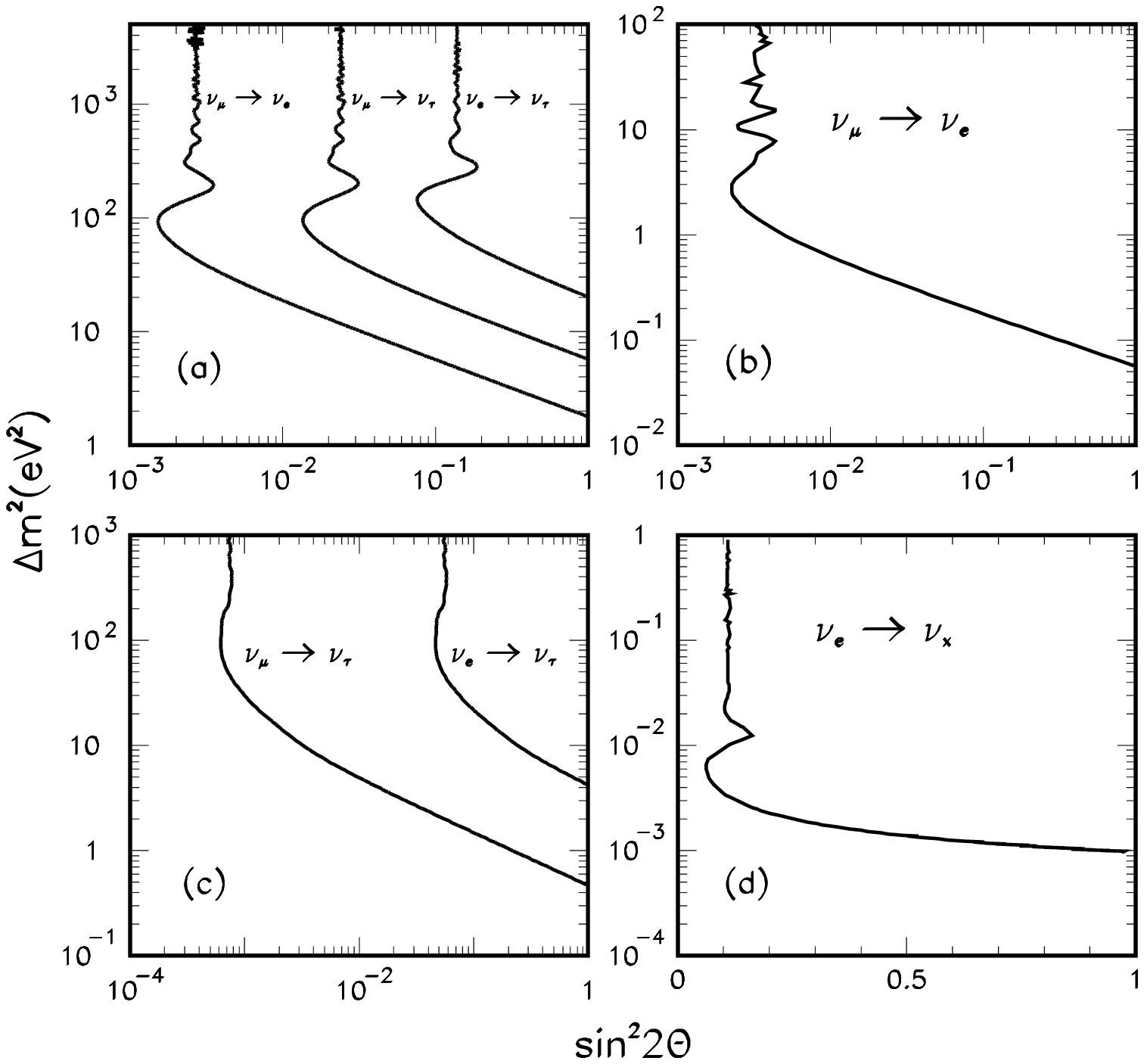,height=20.0cm,width=15.cm}}
\vglue -4.5cm
\caption{ Our reproduction of the limits given by CCFR (a), E776 (b), 
CHORUS (c) and CHOOZ (d) on neutrino oscillations. 
The excluded region  at 90\% C.L. is the one to the right of each curve.}
\label{fig1}
\vglue -1.cm
\end{figure}

In order to extract the CCFR bound, we have performed a statistical 
treatment of the data similar to the one prescribed in Ref.\ \cite{panta}, 
that is, we have minimized the $\chi^2$ function 
 
\begin{equation}
\chi^2 = \sum_{i=1}^{15} \displaystyle \frac{(N_i^{\text{obs}}-N_i^{\text{theo}})^2}{\sigma_i^2},
\label{ccfr1}
\end{equation}
where $N_i^{\text{obs}}$ and $\sigma_i$ are the experimental points and 
errors for the $i$-th bin, respectively, read of from the electron spectrum 
given in Ref.\ \cite{ccfr}. This spectrum has 15 energy bins that go from 
30~GeV to 600~GeV in visible energy and we assume here this to be equal to the
neutrino energy. $N_i^{\text{theo}}$ is the theoretical prediction of the 
 $i$-th bin under a certain oscillation hypothesis. Explicitly, 

\begin{eqnarray}
N_i^{\text{theo}}|_\alpha & = & \langle N^{\text{exp}} \; \{ (P(\nu_e \to \nu_e) 
+ \delta_{\alpha 3} P(\nu_e \to \nu_\tau)\; m(E))   \nonumber \\
                    & + & R(E) [ \delta_{\alpha 2} P(\nu_\mu \to \nu_e) 
+ \delta_{\alpha 1} P(\nu_\mu \to \nu_\tau) \; m(E)] \} \rangle,
\label{ccfr2}
\end{eqnarray}
where $N^{\text{exp}}$ is the expected flux of $\nu_e$ calculated without 
oscillations taken from Ref.\ \cite{ccfr}, $R(E)$ is the ratio between 
the fluxes of $\nu_\mu$ and $\nu_e$~\cite{ccfr},  $m(E)$  is the 
probability of misidentifying $\nu_\tau$ as $\nu_e$ (18 \%) 
times the ratio of $\nu_\tau$ to $\nu_e$ charged current cross 
sections~\cite{x-sec} and $\alpha=1,2,3$ mean respectively 
channels (1), (2) and (3) as explained above. Also, $\langle \rangle$ 
means that we have averaged the 
neutrino propagation length over the neutrino decay pipe and the 
neutrino energy over the bin width. We were able to reproduce the 
experimental exclusion curves for each channel reasonably well. 
We show our results for the three oscillation modes inspected by CCFR in 
Fig.~\ref{fig1}~(a), they can be compared to Fig. 5 of the last 
paper in Ref.\ \cite{ccfr}.

To extract the quantum decoherence limit for each oscillation mode, 
we use the following method~: at fixed $\sin^2 2\theta$  
we minimize the $\chi^2$ given in Eq.~(\ref{ccfr1}) as a function 
of $\gamma_0$, thus obtaining the bounds using 
$\chi^2 - \chi^2_{\text{min}}=2.70$, $3.84$ and $6.63$ at 90~\%, 95~\% and 
99~\%~C.L., respectively. We have verified for each $\sin^2 2 \theta$ 
that $\chi^2_{\text{min}}$ occurs at $\gamma_0=0$.

\subsection{E776}
\label{sec:sec3b}

The E776~\cite{e776} experiment at the Brookhaven National Laboratory 
searched for $\nu_\mu \to \nu_e$ oscillations, looking for $\nu_e$ 
appearance  1 km from the source of a wide-band $\nu_\mu$ beam of average 
energy around 1 GeV. 

The two primary backgrounds for this experiment were $\nu_e$ contamination in 
the beam and $\nu_\mu$-induced $\pi^0$'s which are misidentified as electrons 
by the detector. The E776 data collected was consistent with the expected 
background, so no signal of $\nu_\mu \to \nu_e$ oscillation 
was found~\cite{e776}. 

To reproduce the E776 limits on ${\nu}_e$ ($\bar{\nu}_e$) appearance, we have 
minimized the sum ${\chi}^{2}_{\text{tot}}={\chi}^2 + \overline{{\chi}}^2$,  
which can be derived from a likelihood function and takes into account 
both Poisson statistical distribution of data and the presence of 
background~\cite{bc,fc}, using the definitions 

\begin{equation}
\stackrel{\ufa}{\chi^2}  = 2\sum_{i}{\left[ ( \stackrel{\ufa}{N}
{\opa}_{i}^{\text{th}} + \stackrel{\ufa}{B}{\opa}_{i}
-\stackrel{\ufa}{N}{\opa}_{i}^{\text{obs}}) + \stackrel{\ufa}{N}{\opa}_{i}^{\text{obs}}\ln\left(\frac{\stackrel{\ufa}{N}{\opa}_{i}^{\text{obs}}}{\stackrel{\ufa}{N}{\opa}_{i}^{\text{th}}+\stackrel{\ufa}{B}{\opa}_{i}}\right)\right] ,}
\label{e776-chi}
\end{equation}
where $N_{i}^{\text{obs}}$ ($\overline{N}_{i}^{\text{obs}}$) and 
$N_{i}^{\text{th}}$ ($\overline{N}_{i}^{\text{th}}$) are the observed 
and the theoretical contents of $\nu_e$ ($\overline{\nu}_e$) events in  
the $i$-th bin, in the presence of the background 
level $B_i$ ($\overline{B}_i$). 

$N_{i}^{\text{obs}}$ ($\overline{N}_{i}^{\text{obs}}$) 
and $B_i$ ($\overline{B}_i$) were taken from 
Fig.\ 3 of Ref.\ \cite{e776} and $N_{i}^{\text{th}}$ 
($\overline{N}_{i}^{\text{th}}$) calculated as
\begin{equation}
\stackrel{\ufa}{N}{\opa}_{i}^{\text{th}}=
\int{\stackrel{\ufa}{N}_{{\nu}_{\mu}}(E)P({\nu}_{\mu} \rightarrow {\nu}_{e})
\left(\frac{\stackrel{\ufa}{\sigma}_{{\nu}_{e}}}{\stackrel{\ufa}{\sigma}_{{\nu}_{\mu}}}\right)dE},
\end{equation}
where the integration was performed over the $i$-th bin of
 the $\nu_{\mu}$ ($\overline{\nu}_{\mu}$) spectrum $N_{{\nu}_{\mu}}$
($\bar N_{{\nu}_{\mu}}$), also
given in Ref.~\cite{e776}. The cross sections were taken from 
Ref.~\cite{x-sec}.
We show our exclusion plot for the two generation mass induced oscillation in 
Fig.~\ref{fig1} (b), it agrees well with the E776 plot in Ref.\ \cite{e776}.

To obtain the decoherence limits, we have minimized ${\chi}^{2}_{\text{tot}}$,
and computed the excluded regions for $\gamma_0$ at various confidence 
levels, in the same way as already described for CCFR.

\subsection{CHORUS}
\label{sec:sec3c}

CHORUS~\cite{chorus,chorus1,chorus2,chorus3} experiment at CERN is  
a neutrino oscillation experiment that has been looking for $\nu_\tau$  
appearance in the channels $\nu_\mu \to \nu_\tau$ and $\nu_e \to \nu_\tau$. 
The CHORUS detector is illuminated by a beam consisting 
mainly of $\nu_\mu$ coming from the ``West 
Area Neutrino Facility''. 
The average $\nu_\mu$ momentum is 27 GeV and the average distance 
they have to travel from the target to  CHORUS is about 0.6 km. 
The $\nu_\mu$ beam is also composed of 5\% $\bar \nu_\mu$ and  about 1\% 
$\nu_e+\bar \nu_e$. 
The genuine $\nu_\tau$ content of the beam is estimated to 
be negligible, at the level of $ 3.3 \times 10^{-6}$ $\nu_\tau$ 
charged current interactions per $\nu_\mu$ charged current 
interactions~\cite{chorus2}, well below the sensitivity limit of the 
experiment. The aim is to detect $\nu_\tau$  by observing the 
$\tau^-$ produced in a charged current interaction and its subsequent decay 
vertex in an active nuclear emulsion target. 

The basic difference between the $\nu_\mu \to \nu_\tau$ and 
$\nu_e \to \nu_\tau$ searches comes from the energy spectra, 
the average energy of $\nu_e$ is about 20 GeV higher than $\nu_\mu$, 
which leads to differences in the acceptances for $\nu_\tau$ 
interactions~\cite{chorus3}.  

The main sources of background are the charm $D$ meson production followed 
by  its muonic decay (for the leptonic modes), the  $D$ meson hadronic decays 
and hadronic scattering on target nuclei without any 
visible nuclear breakup or recoil (for the hadronic modes)
and prompt $\nu_\tau$ from the beam~\cite{chorus2,chorus3}.

We define the average oscillation probability
\be
\langle P_{\alpha \to \tau} \rangle = \displaystyle \frac{\int 
\Phi_{\nu_\alpha}(E) P(\nu_\alpha \to \nu_\tau) dE}
{\int \Phi_{\nu_\alpha}(E)dE},
\label{meanprob}
\ee
where $\alpha=\mu,e$, $\Phi_{\nu_\mu}$ and $\Phi_{\nu_e}$ are 
the $\nu_\mu$ and $\nu_e$ fluxes, respectively, taken from 
Ref.\ \cite{chorus} and $P(\nu_\alpha \to \nu_\tau)$ is  the oscillation 
probability given in Eq.~(\ref{simpro}). The exclusion region can be 
calculated according to Refs.~\cite{chorus2,chorus3} by using   
\be
\langle P_{\alpha \to \tau} \rangle \leq  \frac{F}{N^{\text{exp}}_\alpha},
\label{uplim}
\ee
where $N^{\text{exp}}_\alpha$ involves ratios of cross sections, acceptances, 
 efficiencies, number of muonic and non-muonic events and branching ratios 
for the different decay modes~\cite{chorus3} and $F$ is a numerical factor 
for zero $\nu_\tau$ events observed,  no background expected and  a total 
systematic error of 17\% following the prescription of Ref.\ \cite{ch}, 
\ie $F=2.38,3.12$ and $ 4.9$ respectively for 
90\%, 95\%  and 99\% C.L. 

From Ref.~\cite{chorus3}, we can extract:
$N^{\text{exp}}_\mu=5950.0$ for $\nu_\mu \to \nu_\tau$ and 
$N^{\text{exp}}_e=79.3$ for  $\nu_e \to \nu_\tau$.
 Using these experimental inputs in Eq.\ (\ref{uplim}) we were able, 
for $\gamma_0=0$, to reproduce quite well the exclusion regions 
at 90\% C.L. presented by CHORUS. 
This can be seen by comparing our plot for CHORUS in Fig.\ \ref{fig1}(c) with 
Figs.\ 1 and 2 of Ref.\ \cite{chorus3}.
 
The exclusion region for quantum decoherence was obtained by fixing 
$\sin^2 2\theta$ and computing the $\gamma_0$ value which satisfies 
each confidence level requirement given by Eq.~(\ref{uplim}). We did not do 
the analysis for the latest CHORUS result of Ref.\ \cite{chorus4} since we do 
not know exactly what is the total systematic error of this data, hence  
the correct value of $F$ we should use to re-interpret their results. 
In any case, this will not change very much our conclusions. We will comment 
more on this latter on.

\subsection{CHOOZ}
\label{sec:sec3d}

The CHOOZ reactor experiment~\cite{chooz} is the one which provides the 
most restrictive $\nu_e \to \nu_e$  disappearance limit. 
The baseline is about 1 km and the neutrino energy around 3 MeV, so that 
$L/E \sim 300$ m/MeV. Since it is a disappearance experiment, it can be 
employed to infer bounds on $\gamma_0$ in the oscillation modes 
$\nu_e \to \nu_\mu$ and $\nu_e \to \nu_\tau$. 
In order to do this we have used the average of the ratio of the measured 
positron spectrum over the expected one, $R = 1.01 \pm 0.028 (\text{stat}) \pm 0.027(\text{syst}) $~\cite{chooz} 
and compared this with the  $\nu_e \to \nu_e$ survival probability averaged 
over the spectrum. 

We work with the following $\chi^2$ definition
\be
\chi^2 = \left( \frac{\langle 1-P(\nu_e \to \nu_\alpha) \rangle -R}{\sigma_R} \right)^2,
\label{chi2-chooz}
\ee
where $\alpha=\mu$ or $\tau$, $\langle \rangle$ means that the survival 
probability has been averaged in the positron energy spectrum and 
$\sigma_R=0.039$ is the total error in $R$.
We recall that the positron energy and the  neutrino energy are related by 
a simple rescale:  $E_{e^+} = E_{\nu_e} - 1.804$ MeV which allows us to 
easily perform the above mentioned average.
The oscillation probability $P(\nu_e \to \nu_\alpha)$ is given 
in Eq.~(\ref{simpro}). Again, the excluded region is computed using the 
method already described in Sec.~\ref{sec:sec3a}. 
We have adopted this elementary $\chi^2$ formula since it allows us 
to reproduce almost exactly the exclusion curve obtained by the analysis 
$A$ of Ref.\ \cite{chooz}. This can be seen in Fig.~\ref{fig1} (d).
We have checked that the global minimum of the $\chi^2$ always occurs 
at $\gamma_0=0$.

\section{Constraints from Current Experimental Data} 
\label{sec:sec4}

Here, we present the limits on $\gamma_0$ given by experimental data in 
 three different neutrino oscillation modes. We display in 
Figs.\ \ref{fig2}-\ref{fig4} our exclusion plots in the 
plane $\sin^2 2 \theta \times \gamma_0$ at 99\% C.L., 
they give a general view of our results and show the 
sensitivity of each experiment for each oscillation mode.

Experiments with negative results on neutrino oscillations, such as the 
ones studied here,  have a natural restriction to impose limits 
on $\gamma_0$. This happens when 
$\sin^2 2 \theta \to 2 P_{\text{min}}$. At this point the right hand side 
of Eq.\ (\ref{pmaxlim}) goes to 1, then for any value of $\gamma_0$
the left hand side (that can only go from 0 ($\gamma_0 \to 0$) to 
1 ($\gamma_0 \to \infty$)) satisfies the inequality, so that 
we are not sensitive to variations of $\gamma_0$. 
Because of this fact, we cannot explore the SA case 
for $\nu_e \to \nu_\tau$, or use CHOOZ, instead 
of E776, to put restrictions on $\gamma_0$ in the mode
$\nu_\mu \to \nu_e$ with $n=-1$ in the SA case. These features can be 
seen in Figs.~\ref{fig3} and \ref{fig4}.

In the following, we will discuss the regions of special interest, in 
each mode, as  underlined in Sec. \ref{sec:sec3}.

\subsection{$\nu_\mu \to \nu_\tau$}
\label{sub1}

\begin{figure}
\vglue -4.0cm
\centerline{
\epsfig{file=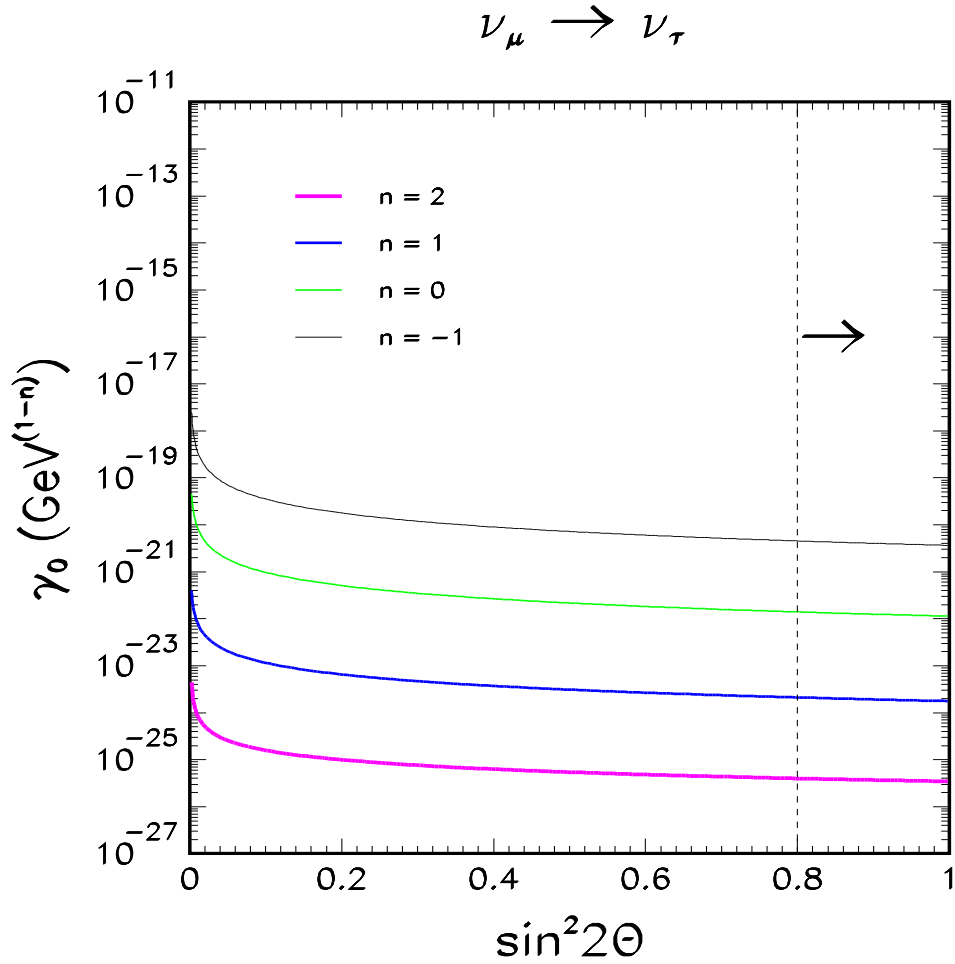,height=20.0cm,width=15.8cm}}
\vglue -4.8cm
\caption{ Limits on $\gamma_0$ as a function of $\sin^2 2\theta$ in the  
$\nu_\mu \to \nu_\tau$ mode, for  
 $n=-1,0,1$ and  $2$. The excluded region at 
99 \% C.L. is the one to the right of each curve.
The lower value of the amplitude consistent with the ANP is marked by 
a dashed line with an arrow.
All the limits were obtained with CHORUS data.}
\label{fig2}
\vglue -1.cm
\end{figure}

The best limits in this oscillation mode for all values of $n$ are given by 
the CHORUS experiment, see Fig.~\ref{fig2}. 

Previous to comment our bounds in this channel, we will make a 
brief remark about the constraints on $\gamma_0$ from atmospheric
neutrino experiments, since they seem to imply
$\nu_\mu \to \nu_\tau$ oscillations\ \cite{sterile}. These 
experiments are clearly in 
great advantage over terrestrial ones, 
they cover oscillation baselines from approximately 500 km to 
about 12\, 000 km, the diameter of the Earth, with neutrino energies 
from $\sim$ 1 GeV up to a few 
hundred GeV. One can estimate the order of magnitude 
of their  bounds on $\gamma_0$ by asking 
 ${\cal O} (2\gamma L) \lsim 1$ using $L \sim 10^{4}$ km 
and $ <E_\nu> \sim 10^2$ GeV (the most energetic 
upward going neutrinos will push the limit), this gives : 
$\gamma_0 \lsim 10^{-21}$ GeV$^{2}$ for $n=-1$,
$\gamma_0 \lsim 10^{-23}$ GeV for $n=0$,
$\gamma_0 \lsim 10^{-25}$ for $n=1$ and 
$\gamma_0 \lsim 10^{-27}$ GeV$^{-1}$ for $n=2$. This is in good agreement 
with the limits obtained in Ref.\ \cite{lisi}~\footnote{We remark that in our 
notation $2 \gamma_0$ corresponds to $\gamma_0$ of Ref.\ \cite{lisi}.} 
for $n=-1,0$ and $2$  using a statistically rigorous analysis of 
the atmospheric neutrino data.

We display the bounds on $\gamma_0$ as a function of $\sin^2 2 \theta$  
in Fig.\ \ref{fig2}, for $n=-1,0,1$ and $2$ at 99\% C.L. 
In this figure the lower limit of the amplitude region compatible with the 
standard solution to the ANP is marked by a line with an arrow.  
For $n=2$, the analysis of the CCFR data described in 
Sec.\ \ref{sec:sec3a} also can provide a very similar limit, slightly less 
restrictive than the one given by CHORUS. We do not show this here. 

We observe that some of our bounds are substantially weaker than the 
ones given by the atmospheric data. Notwithstanding, they are independent 
limits and the most restrictive ones that can be computed with data from 
neutrinos produced in accelerators up to this day.

Finally, it is easy to re-scale our results to obtain more stringent
constraints for the combined CHORUS/NOMAD 
actual data analysis~\cite{chorus-nomad}. 
In fact, it is possible that in  the near future 
CHORUS and NOMAD combined analysis may even provide a better bound than the 
atmospheric neutrino experiments for the case of $n=-1$. However,  
it seems unlikely that they will be able to overcome the atmospheric limit 
on $\gamma_0$ for $n=0,1$ and $2$.

\subsection{$\nu_\mu \to \nu_e$}
\label{sub2}
\begin{figure}
\vglue -4.0cm
\centerline{
\epsfig{file=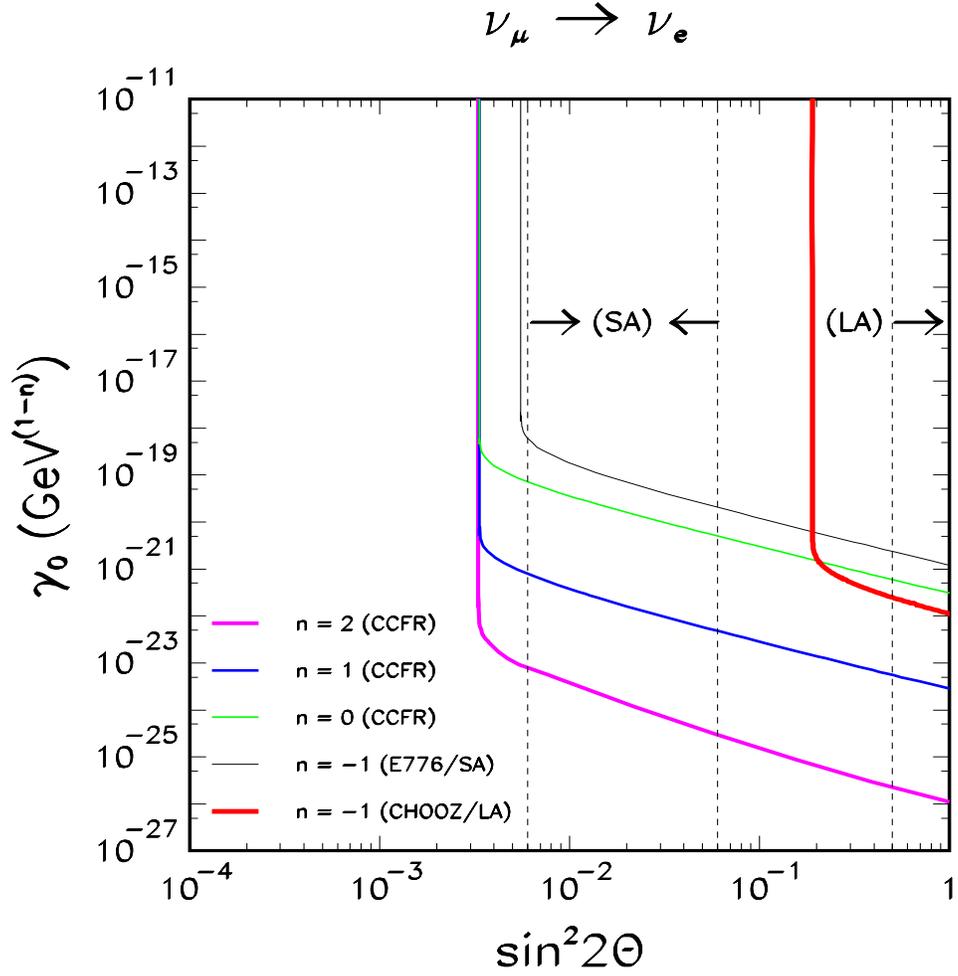,height=20.0cm,width=15.8cm}}
\vglue -4.5cm
\caption{ Limits on $\gamma_0$ as a function of $\sin^2 2\theta$ in the 
$\nu_\mu \to \nu_e$ mode, for $n=-1$ (SA, LA), $0,1$ and $2$ 
at 99~\% C.L.
}
\label{fig3}
\vglue -1.cm
\end{figure}

\begin{figure}
\vglue -4.0cm
\centerline{
\epsfig{file=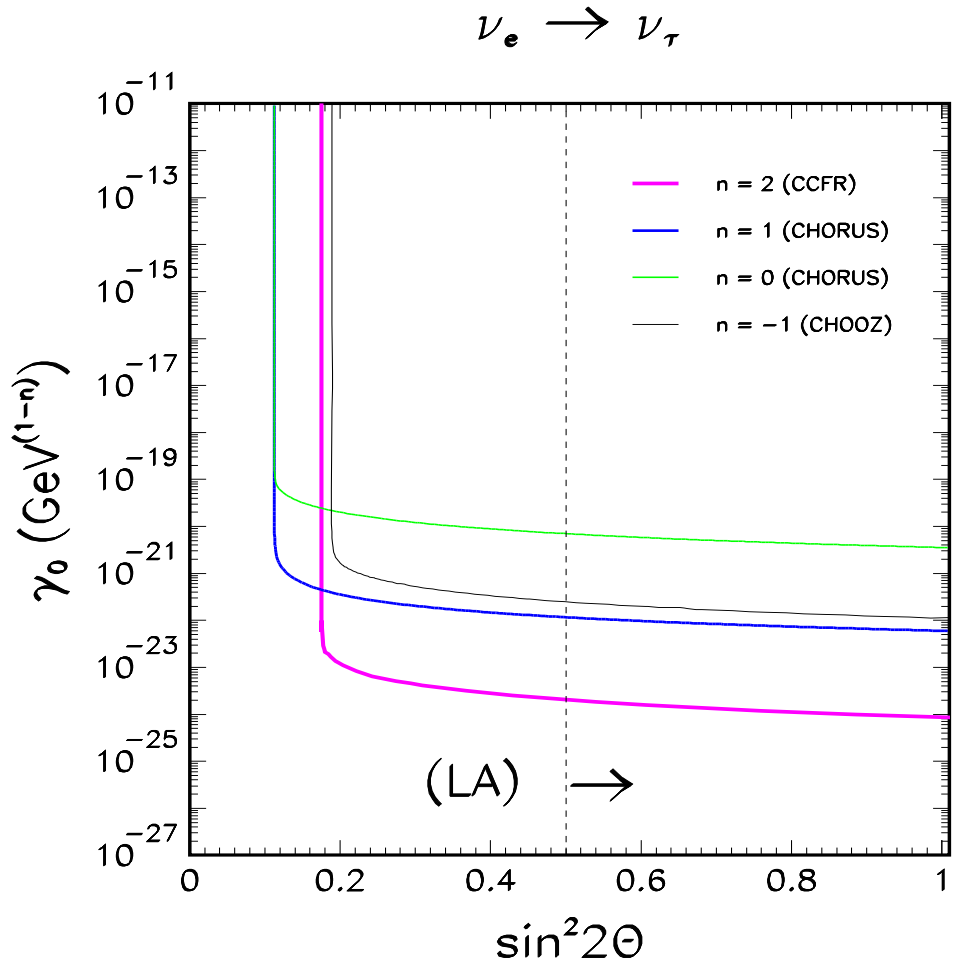,height=20.0cm,width=15.8cm}}
\vglue -4.8cm
\caption{ Limits on $\gamma_0$ as a function of $\sin^2 2\theta$ in the $\nu_e \to \nu_\tau$ mode, for $n=-1,0,1$ and $2$ at 99~\% C.L. 
}
\label{fig4}
\vglue -0.05cm
\end{figure}

The CHOOZ/E776 experiments provide the best limits for $n=-1$ and 
CCFR  for $n=0,1$ and $2$. This can be seen in Fig.~\ref{fig3}. 
Here, we focus on two extreme situations: LA and SA.

Before discussing our bounds in this channel, we 
need to introduce a comment linking the SNP and quantum decoherence. 
This is in order to make a comparison between our results and the 
possible ones in the Sun. 

In the case of solar neutrinos, 
we can envisage two different approaches: (i) decoherence influences 
long wavelength oscillations or (ii) decoherence perturbes MSW-type 
oscillations. In the event of (i), $L$ has to be taken as the Sun-Earth 
distance, $L \sim 1.5 \times 10^{8}$ km, in the event of (ii) as the Sun 
radius, $L_{\odot} \sim 7 \times 10^{5}$ km. Using the fact 
that, the quantum decoherence parameter in order 
to be revealed must satisfy ${\cal O} (2\gamma L) \sim 1$, and 
the range  
of neutrino energy is $10^{-4} \text{ GeV} \lsim  E_\nu 
\lsim 10^{-2} \text{ GeV}$, we can estimate the order of 
$\gamma_0$ for both situations. Considering (i), we get  
$\gamma_0 \sim 7 \times \{ (10^{-32}-10^{-30})\, \text{GeV}^2, 10^{-28} \, \text{GeV},10^{-26}-10^{-24},(10^{-24}-10^{-22})\, \text{GeV}^{-1} \}$  and  
considering (ii) 
 $\gamma_0 \sim 1.4 \times \{(10^{-29}-10^{-27})\,\text{GeV}^2, 10^{-25} \, \text{GeV},10^{-23}-10^{-21},(10^{-21}-10^{-19})\, \text{GeV}^{-1} \}$, respectively for $n=-1,0,1,2$. 
This means that the solar neutrino data can at the most set bounds 
on $\gamma_0$ within the ranges quoted above.

In the case of LA, we show our results in Fig. \ref{fig3} for
$n=-1,0,1$ and $2$ at 99\% C.L., we see for $n=-1,0$ that 
 the CHOOZ and CCFR limits are weaker than the possible constraints one 
can get from solar neutrinos. 

For $n=1$, the CCFR constraints 
are in the fringe of the solar sensitivity for long wavelength oscillations.
Nevertheless, given that our 
results at $\sin^2 2 \theta=1$
can also be understood 
as limits on the PDM alone, we can conclude that, for $n=1$ 
the possibility of explaining the solar neutrino 
data by this mechanism is in fact 
discarded. This is because not even the total rate measured by  
Super-Kamiokande can be explained if 
$\gamma_0 \lsim  3 \times 10^{-24}$,  
unless we admit that the $^8$B  flux can be 
50~\% smaller, \ie 3 sigma away from the central value of the Solar 
Standard Model prediction. For MSW-type oscillations and $n=1$, 
the CCFR limits are clearly more 
restrictive than those that could be reached by solar neutrino data. 
Obviously, the solar neutrino problem cannot be solved by 
the PDM in this case. 

When $n=2$, the CCFR limits
are much lower than the solar sensitivity 
for both (i) and (ii). Here again, the PDM cannot be 
solution of the solar neutrino problem.

The general result is that for $n \gsim 2$ (i) and $n \gsim 1$ (ii), one 
cannot hope to improve the CCFR bounds with solar data nor to explain 
the SNP by the PDM. This is due to the energy dependence of the 
limit on $\gamma_0$ (see Eq.\ (\ref{gam})) 
and the fact that CCFR average neutrino energy is much higher
than that of solar neutrinos. Therefore, the tendency 
is that the CCFR limits become stronger than the solar 
neutrino ones, with increasing $n$.

For SA, we can also see the limiting curves in 
Fig. \ref{fig3}. Note that, in this case, the constraints for $n=-1$ 
were derived using E776 data.

For the case $n=-1,0$ and $1$, we expect 
that the bounds with solar neutrinos 
should be much better than the ones we present here. 
In the case $n=2$, the CCFR limits are 
below the solar sensitivity only for the situation (ii).  

As mentioned in Sec.~\ref{sec:sec3}, the SA bound for $\nu_\mu \to \nu_e$ 
mode can be kept also valid under the hypothesis that the 
oscillation parameters lay in the 
LSND/\-KAR\-MEN/\-Bu\-gey~\cite{lsnd,karmen,bugey}
allowed region ($3 \times 10^{-3} \leq  \sin^2 2\theta \leq 3 \times 10^{-2}$, $0.2 \text{ eV}^2 \lsim \Delta m^2 \lsim 2 \text{ eV}^2$). 
Due to this range of $\Delta m^2 $, just the bounds with CCFR 
data ($n=0,1,2$) can be used if the LSND mass scale is adopted. 

To  compare our constraints with the possible ones that could be extracted 
from the LSND data, 
we have to do some estimations. Now, working with the 
assumption   
${\cal O} (2\gamma L) \sim 1$ applied to LSND using $< E_\nu> = 40$ MeV 
and $L$ = 30 m, we arrive at 
$\gamma_0\sim \{ 10^{-19}\, \text{GeV}^2, 3\times 10^{-18} \, \text{GeV}, 
0.8 \times 10^{-16}, 2 \times 10^{-15} \, \text{GeV}^{-1} \}$ respectively 
for $n=-1,0,1,2$.  This means that, for $n=-1$,  we are limiting 
the same range of $\gamma_0$ that could be achieved with the LSND data.
Meanwhile, for all the others $n$ the CCFR limits are much below 
the LSND sensitivity. Furthermore, if we compare 
these sensitivities with our bounds on the  PDM ($\sin^2 2\theta=1$), it is 
evident that the LSND results cannot be explained solely by decoherence.

\subsection{$\nu_e \to \nu_\tau$}
\label{sub3}
In this mode, the CHOOZ experiment will provide the best constraints 
when $n=-1$, CHORUS when $n=0,1$ and CCFR when $n=2$, as can be seen in 
Fig.~\ref{fig4}.  

As we already mentioned in the beginning of this section, here we can only 
discuss the LA case. 
We verify in this oscillation mode that only for $n \gsim 2$ the 
experimental limits imposed by CCFR are stiffer than 
those that could be achieved by the solar neutrino data.

\section{Future Perspectives} 
\label{sec:sec5}

In this section, we briefly discuss the perspectives to 
improve the bounds presented before, using the mean values of $E_\nu$, $L$ 
and the sensitivities  expected for the forthcoming neutrino oscillation 
experiments such as KamLAND~\cite{kamland}, MINOS~\cite{minos,minostec,numi}, 
the CERN-to-Gran Sasso neutrino experiments ICANOE/OPERA~\cite{ngs,icanoe} 
and a possible neutrino factory in a muon collider~\cite{geer}.
 
To estimate the limits on $\gamma_0$ that could be reached at prospective 
long baseline facilities, we investigate in turn the two possible outcomes 
of the experiments : (i) no signal and (ii) a positive signal of 
neutrino appearance/disappearance is observed.

Experiments that give negative results are insensitive to the mass scale, 
so when considering (i) we can work with Eq.\ (\ref{pmaxlim}), replacing 
$\sin^2 2\theta$ by 1, and the mean value of $E_\nu$ and $L$
as well as the sensitivity goals by the corresponding ones of each 
proposed experiment. In this manner, at 
this fixed value of $\sin^2 2\theta$, we can test simultaneously 
decoherence plus mass oscillation and the PDM alone. This is done 
for $\nu_\mu \to \nu_e$ and $\nu_e \to \nu_\tau$ 
modes only, $\nu_\mu \to \nu_\tau$ is assumed to be already observed by 
Super-Kamiokande and K2K. This will be a rough estimate since we completely 
neglect matter effects, which could be crucial in determining the exact 
sensitivity of each experiment to $\gamma_0$. Nevertheless, we believe that 
this will affect only the actual value of the limit but will not change 
our conclusions on which will be the most restricting experiment at each $n$ 
and oscillation mode.  

\begin{figure}
\vglue -4.cm
\centerline{
\epsfig{file=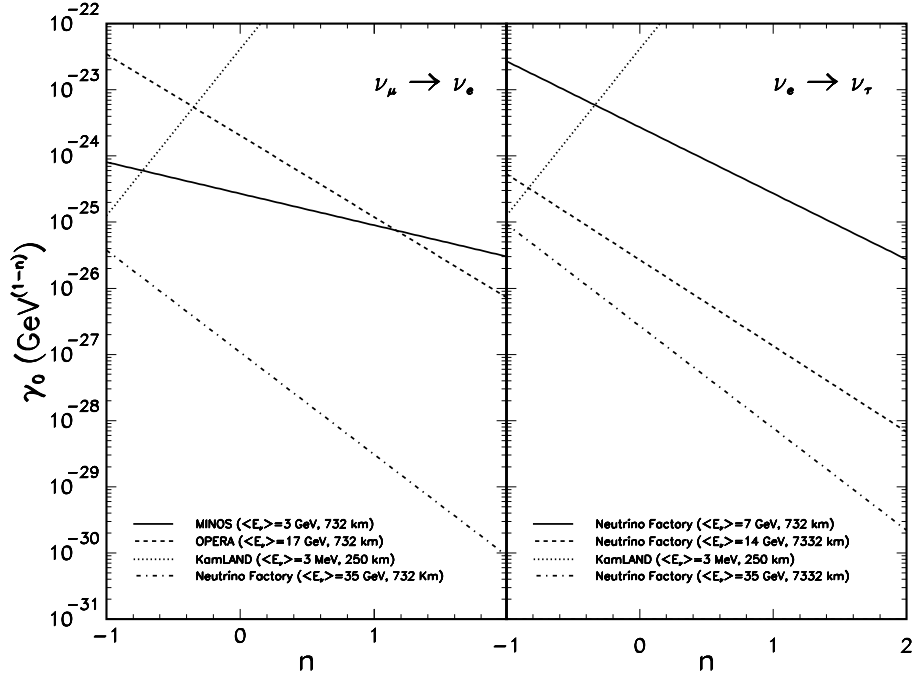,height=9.5cm,width=13.3cm}}
\vglue 0.5cm
\caption{Limits on $\gamma_0$ for some of the next generation neutrino 
long baseline experiments assuming the proposed sensitivity 
of each experiment and no signal of oscillation observed in 
the $\nu_\mu \to \nu_e$ and $\nu_e \to \nu_\tau$ modes.
}
\label{fig6}
\vglue -0.05cm
\end{figure}

In Fig.\ \ref{fig6}, we plot our limit on $\gamma_0$  
in the $\nu_\mu \to \nu_e$ and $\nu_e \to \nu_\tau$ oscillation channels
as a function of $n$ for a few experimental configurations in case 
that no oscillation is observed. 

Let us discuss the $\nu_\mu \to \nu_e$ 
mode first. 
For $n=-1$, KamLAND can bring the current bound three 
orders down, as well as certain configurations of a neutrino factory in a 
muon collider as exemplified in the plot. For $n=0$, MINOS/ICANOE, since 
they have the same $L$ and proposed sensitivity, can improve the limit by 
three orders of magnitude. OPERA will be less powerful since its 
proposed sensitivity is somewhat lower. A neutrino factory can bring this 
limit almost six orders of magnitude down. 
For $n=1$, MINOS/OPERA/ICANOE 
can lower the bound by about a factor ten, but the neutrino factory can push 
it close to $\gamma_0 \lsim 10^{-29}$. In the case $n=2$, the CCFR 
constraint can only be overcome by a future neutrino factory. 

In the $\nu_e \to \nu_\tau$ mode for $n=-1$, KamLAND can certainly bring the 
CHOOZ limit down by a few orders of magnitude. On the other 
hand, some of the proposed neutrino 
factories can get better limits 
on $\gamma_0$, for all the 
values of $n$, than the ones discussed in this paper. 

\begin{figure}
\vglue -4.0cm
\centerline{
\epsfig{file=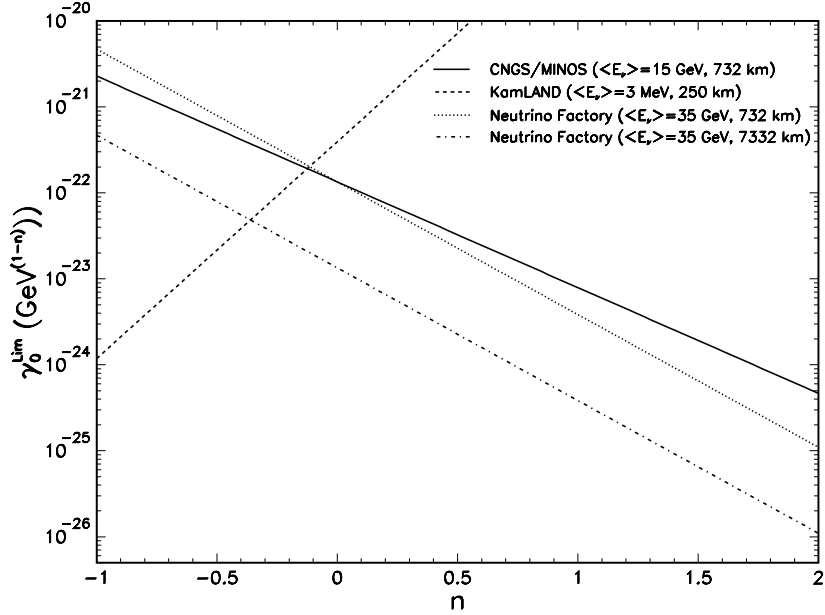,height=8.64cm,width=12.cm}}
\vglue 0.5cm
\caption{Attainable limits on $\gamma_0$ for some of the next generation 
neutrino long baseline experiments assuming they will observe neutrino 
oscillation.
}
\label{fig7}
\vglue -0.05cm
\end{figure}

When considering (ii) to put a bound on $\gamma_0$, the oscillation 
probability must be affected by the decoherence parameter so that 
${\cal O} (2 \gamma L) \sim 1$. Thus, this value will not depend 
on the mode of oscillation, but only on the characteristic $L$ and 
mean $E_\nu$  of each experiment.
This gives us an idea of their sensitivity to $\gamma_0$ as a 
function of $n$ which is shown in Fig.\ \ref{fig7}, for a few 
situations. 

For $\nu_\mu \to \nu_\tau$ only the {\em ansatz} $n=-1$ 
can, in principle, be better tested in a neutrino factory than 
by using the atmospheric neutrino data. This will greatly 
depend on the specific choice of baseline and of the muon energy. 

In the mode $\nu_\mu \to \nu_e$, the bound on $\gamma_0$ with the 
{\em ansatz} $n=-1$ can be improved by KamLAND. For $n \gsim 2$, the CCFR 
limits we give in this paper can be lowered by some of the experiments we 
studied.

In the mode $\nu_e \to \nu_\tau$, KamLAND ($n=-1$), 
MI\-NOS\-/I\-CA\-NOE/O\-PE\-RA/neu\-tri\-no factory ($n=0,1$) and a 
neutrino factory ($n=2$) can improve the present limits, 
but only by one to two orders of magnitude.

\section{Conclusions} 
\label{sec:sec6}

We have analyzed the experimental data from the terrestrial neutrino 
oscillation experiments CHOOZ, CHORUS, E776 and CCFR in order to extract 
constraints on the decoherence parameters  in the framework of two 
generation neutrino oscillations induced by mass plus quantum 
dissipation. This was done for the so called weak limit, where 
all decoherence parameters, except for $\gamma$, are supposed to 
be either zero or out of experimental reach.  
We have derived constraints on $\gamma_0$ under four different 
{\em ansatz}, \ie $\gamma = \gamma_0 \, E_{\nu}^{n}$,  with $n=-1,0,1,2$,  
from examining experimental data in the channels 
$\nu_\mu \to \nu_\tau$, $\nu_\mu \to \nu_e$ 
and $\nu_e \to \nu_\tau$. This is valid from  
distinct upper values of $\Delta m^2$, depending 
on each experiment so that we can always consider $\Delta m^2$ negligible 
in the probability, down to $\Delta m^2=0$.
In particular, we discuss our limits 
paying attention to the phase space of 
($\Delta m^2$, $\sin^2 2\theta$) consistent with 
the standard solution to the ANP for $\nu_\mu \to \nu_\tau$, and
to the SNP in the channels $\nu_e \to \nu_\mu$~(LA, SA) and 
$\nu_e \to \nu_\tau$~(LA). Our limits on $\gamma_0$ for $\nu_\mu \to \nu_e$ 
(SA) are also valid for $\Delta m^2$ in the LSND allowed region.
In addition, at $\sin^2 2\theta=1$, our bounds 
can be also read as direct limits on the PDM. At the end, we have also 
discussed the perspectives of future experiments to push our 
limits further down.

In the $\nu_\mu \to \nu_\tau$ mode (ANP solution range), we have established 
the following bounds: $\gamma_0<(5.6-4.3) \times 10^{-21}$ GeV$^2$ for $n=-1$,
$\gamma_0<(1.6-1.2) \times 10^{-22}$ GeV for $n=0$,
$\gamma_0<(3.2-2.4) \times 10^{-24}$  for $n=1$ and 
$\gamma_0<(4.0-3.1) \times 10^{-26}$ GeV$^{-1}$ for $n=2$, at 99\% C.L.
In spite of the fact that these limits are much less restrictive than the 
ones given in Ref.\ \cite{lisi} from analyzing atmospheric neutrinos, they 
are valuable to be known, since they are independent constraints.

In the $\nu_\mu \to \nu_e$ mode (LA), we have established the 
following bounds: $\gamma_0<(2.5-1.2) \times 10^{-22}$ GeV$^2$ for $n=-1$,
$\gamma_0<(6.0-3.1) \times 10^{-22}$ GeV for $n=0$,
$\gamma_0<(5.5-3.0) \times 10^{-24}$ for $n=1$ and 
$\gamma_0<(2.2-1.2) \times 10^{-26}$ GeV$^{-1}$ for $n=2$,  at 99\% C.L.
From these constraints, we concluded that for $n \gsim 1$  one is discouraged  
to try to extract better limits from the solar neutrino data itself. 
Moreover, these constraints exclude any possibility to explain the 
LSND results by the PDM alone. 

In the $\nu_\mu \to \nu_e$ mode (SA), we have established the 
following limits: $\gamma_0<(6.0-0.27) \times 10^{-19}$ GeV$^2$ for $n=-1$,
$\gamma_0<(7.0-0.6) \times 10^{-20}$ GeV for $n=0$,
$\gamma_0<(7.0-0.5) \times 10^{-22}$ for $n=1$ and 
$\gamma_0<(8.0-0.3) \times 10^{-24}$ GeV$^{-1}$ for $n=2$, at 99\% C.L.
In the case $n \gsim 2$, the solar neutrino data will give weaker
bounds than ours. Besides, for $n \gsim 0$, our constraints 
are stronger than what we could obtain with LSND data.

In the $\nu_e \to \nu_\tau$ mode (LA), we have established the 
following limits: $\gamma_0<(2.5-1.1) \times 10^{-22}$ GeV$^2$ for $n=-1$,
$\gamma_0<(1.0-0.5) \times 10^{-20}$ GeV for $n=0$,
$\gamma_0<(1.3-0.7) \times 10^{-22}$ for $n=1$ and 
$\gamma_0< (2.0-1.0) \times 10^{-24}$ GeV$^{-1}$ for $n=2$ at 99\% C.L.
Here again, for $n \gsim 2$, our bounds 
are stronger than the solar ones.

In the future, there will be many experiments capable of 
further constraining $\gamma_0$. However, it is rather difficult to say 
something completely definite about this since these experiments may or 
may not observe a positive signal of oscillation. In general, we can say that 
in the $\nu_\mu \to \nu_\tau$ mode only for the {\em ansatz} $n=-1$ the current 
limits from atmospheric data, can be improved by a future neutrino factory; in 
the $\nu_\mu \to \nu_e$ and $\nu_e \to \nu_\tau$ channels KamLAND ($n=-1$) 
and a future neutrino factory ($n=0,1,2$) are 
the best candidates to test decoherence. We point 
out that in $\nu_\mu \to \nu_e$ mode (LA), for 
$n \gsim 2$, the constraints discussed here can hardly be overcomed by 
any prospective experiment that observes a positive signal of oscillation. 
We remark that for $n \lsim 1$ the solar neutrino 
data is certainly the best probe for quantum decoherence in 
$\nu_\mu \to \nu_e$ and $\nu_e \to \nu_\tau$ modes.

Recently, bounds on pure quantum decoherence have been determined in 
Ref.\ \cite{kps} from SN1987A data. In fact, using our assumptions on the 
$\gamma$ energy behavior and  the results of Ref.\ \cite{ssb}, we can 
estimate that SN1987A can provide the following limits, 
using $L \sim 50$ kpc and $P(\nu_e \to \nu_\mu,\nu_\tau)< $ 0.14 
for $\langle E_\nu \rangle =26 $ 
MeV~\cite{ssb}:
$\gamma_0 < 0.6 \times 10^{-39}$ GeV$^2$ for $n=-1$, 
$\gamma_0 < 0.2 \times 10^{-37}$ GeV for $n=0$, 
$\gamma_0 < 0.8 \times 10^{-38}$ for $n=1$ and
$\gamma_0 < 0.3 \times 10^{-34}$ GeV$^{-1}$ for $n=2$. 
It is very important to note that, these limits are not comparable to our 
bounds. This is because we are
 assuming decoherence plus mass induced oscillation with  
 $10^{-11} \text{ eV}^2 \lsim \Delta m^2 \lsim 2 \text{ eV}^2$. In this 
 context, the oscillation probability for large 
 propagation distances will be no different than the one for mass induced 
 oscillation, \ie $P(\nu_\alpha \to \nu_\beta) \to \sin^2 2\theta/2$ (see 
Eq.~(\ref{proba})), therefore independent of $\gamma$. Consequently, 
 supernovae or other far away astrophysical objects emitting neutrinos 
 cannot provide limits on  the quantum decoherence parameter $\gamma_0$ 
 in our framework.

On the other hand, even in the situation that one can compare the 
 SN1987A limits with the ones obtained here, \ie for the PDM 
 $(\sin^2 2\theta =1)$, the constraints extracted from reactor and 
 accelerators experiments, where the neutrino fluxes are well controlled, 
  are very robust and worthwhile to be known.

\acknowledgments

We thank GEFAN for valuable discussions 
and useful comments. This work was supported by Conselho Nacional de 
Desenvolvimento Cient\'{\i}fico e Tecnol\'ogico (CNPq) and by Funda\c{c}\~ao 
de Amparo \`a Pesquisa do Estado de S\~ao Paulo (FAPESP).


\newpage



%
%

\newpage

%
%

\newpage

%
%

\newpage

%
%

\clearpage

%
%

%
%



\end{document}